\documentclass[a4paper]{article}
\usepackage{epsfig,amsmath,amsfonts,amssymb,setspace,multirow,textcomp}
\usepackage[T1]{fontenc}
\makeatletter
\renewcommand{\@evenfoot}{\hfil \thepage \hfil}
\renewcommand{\@oddfoot}{\hfil \thepage \hfil}
\makeatother

\renewenvironment{thebibliography}[1]{\begin{oldthebibliography}{#1}\setlength{\parskip}{0ex}\setlength{\itemsep}{0ex}}{\end{oldthebibliography}}
\begin{document}
\fontsize{11}{11}\selectfont % the font size cannot be changed in any case!
%  insert your title, authors information and text instead of the one provided below
%{\text {UDK 523.68}}
\title{Optical Flare Search on the RS CVn-type flare stars AR Lacertae}
\author{Ya.\,S.~Markus, B.\,E.~Zhilyaev}

\date{\vspace*{-6ex}}
\maketitle
\begin{center} {\small $Main Astronomical Observatory, NAS \,\, of Ukraine, Zabalotnoho \,27, 03680, Kyiv, Ukraine$}\\
%%$^{2}International Center for Astronomical, Medical, and Ecological Research, NAS \,\, of Ukraine, 03680, Kyiv, %%Ukraine$}\\
%\begin{center} {\small $International Center for Astronomical, Medical, and Ecological Research, NAS \,\, of %Ukraine, Zabalotnoho \,27, 03680, Kyiv, Ukraine$}\\
{\tt markusyana@gmail.com}
\end{center}

\begin{abstract}
%The abstract of your paper should be here.\\[1ex]
We present the results of fast spectrophotometry of flares on the RS CVn-type star AR Lac with a time resolution of 34 s and a spectroscopic resolution R $\sim $ 1300. The observations were performed on July 21-22, 2021 with the 2.0 m Karl Zeiss telescope at the Terskol Observatory.  During the flares, an additional emission appeared in the spectrum of AR Lac at wavelengths nearby Ca II H\&K ($\lambda $ = 3933, 3968 \AA), nearby Si IV lines ($\lambda $ = 4089, 4116 \AA).  Variations in these lines range from 2 to 5\%. We have estimated the UBV magnitudes from spectrograms through a mathematical convolution of the spectra with the filter transmission curves. The flare amplitudes in the UBV band up to 0.8 magnitudes were discovered. A detailed colorimetric analysis has allowed important parameters of the flares on AR Lac to be estimated: the temperatures at maximum light and their sizes. The color-color $(U-B)$ - $(B-V)$ diagrams confirm that all the flares at maximum light radiate as a blackbody. The temperatures at maximum light were up to 12000 $\pm $ 300 K.  Based on our colorimetric analysis, we have estimated the linear sizes of the flares at maximum light. The linear size of the flares at the maximum luminosity is approximately 2\% of the radius of the star.

{\bf Key words:}\,\,methods: observational; stars: individual: AR Lacertae; techniques: spectroscopy

\end{abstract}

%\section*{\sc introduction}
\section*{\sc introduction}
\indent \indent As presented in Berdyugina's review \cite{Berdyugina}, RS CVn stars represent a class of close detached binaries with the more massive primary component being a G-K giant or subgiant and the secondary a subgiant or dwarf of spectral classes G to M.
They show optical variability interpreted as the rotationally modulated effect of cool spots on their surfaces. The primary appears more active than the secondary. Since they are tidally locked close binaries, they are also fast rotators.

Thus, similar to other cool active stars, RS CVn-type variables are remarkable due to strong chromospheric plages, coronal X-ray, and microwave emissions, as well as strong flares in the optical, UV, radio, and X-ray.
Large amplitude brightness variations of RS CVn stars imply the presence of enormous starspots on their surfaces covering up to 50\% of the visible disc. Remarkable activity and high luminosity of these stars make them
favourite targets for light curve modelling, Doppler imaging and spectral line analysis. Most of the present knowledge on starspots is based on studies of this type stars.

The totally eclipsing binary AR Lac (HD 210334, BD+4 53813) is a well known RS CVn double-lined system whose components, a G2 IV and a K0 IV star, show signatures of an in tense solar-like activity (see eg. Lanza et al. \cite{Lanza} and references therein).

The light curve solution made by Lanza et al. \cite{Lanza} supports  the existence of compact starspots on the G2 IV primary star.

Several flare events have been observed on AR Lac in various spectral regions from radio \cite{Rodono1} to X-ray \cite{Ottmann}, \cite{Rodono2} including UV and EUV chromospheric and transition region lines \cite{Neff}, \cite{Walter}. The flare time-scales generally range from a few hours to a few tens of hours. The long-lasting X-ray flare observed by Ottmann \& Schmitt \cite{Ottmann} had a total duration of about 20h, including two small events during the decay phase. Moreover, the typical flare light curve was followed by along-duration enhanced emission, lasting more than one orbital period.

A high-cadence search was conducted on the known RS CVn-type flare stars AR Lac, II Peg, and UX Ari. Two optical flares were observed in the B-band on AR Lac at 5 milliseconds (ms) resolution for a rate of 0.04 flare/hour. \cite{Haagen}.

The flare time-scales generally range from a few hours to a few tens of hours. In our work we present results of optical flare search on AR Lac on  short time-scales. We demonstrate flare events  on  time-scales of minutes.

\section*{\sc Observations and data processing}
%\section*{\sc OBSERVATIONS AND DATA PROCESSING}% \label{sec:style}

We collected a total of 198 spectra of the RS CVn star AR Lac on July 21-22, 2021 with the 2.0 m Karl Zeiss telescope at the Terskol Observatory. The observations were obtained at the Cassegrain spectrograph with intermediate resolution 1300 over the wavelength range 3500 - 5800 \AA.

The flat-field spectra were obtained and subsequently averaged. We obtained the short "dark" spectra after each exposure as well as "sky" spectra for subsequent processing. Pixel-to-pixel sensitivity variations were removed by dividing the stellar spectra by a mean flat-field spectrum assembled. 

All the spectra collected were processed uniformly by means of programs in Interactive Data Language (IDL). The baseline response of the CCD detector was subtracted from all the spectra by the acquisition software at the Terskol Observatory.

To provide wavelength calibrations, we obtained spectra of the sky at the end of a night. The final spectral time series is represented by a two-dimensional array in-plane wavelength - time (Fig. 1). 

Fig. 2 shows the white light curve of AR Lac as a wavelength-coordinate sum. One has clearly seen flare events lasting from half a minute to five minutes for a rate of 6 flare/hour on an average.

In Fig. 4 we can see the peaks of activity at wavelengths nearby Ca II H\&K ($\lambda$ = 3933, 3968 \AA), nearby Si IV lines ($\lambda$ = 4089, 4116 \AA). The relative variations are equal to the ratio between the rms deviations in the array of spectra and the average intensity of the spectrum. Variations in these lines range from 2 to 5\%.

It is important to note the presence of activity in silicon Si IV lines ($\lambda$ = 4089, 4116 \AA). This indicates the presence of hot plasma with a temperature above 10,000 K associated with flares on the surface of a star.

%+++
\begin{figure}[!h]
\centering
\begin{minipage}[t]{.45\linewidth}
\centering
\epsfig{file = 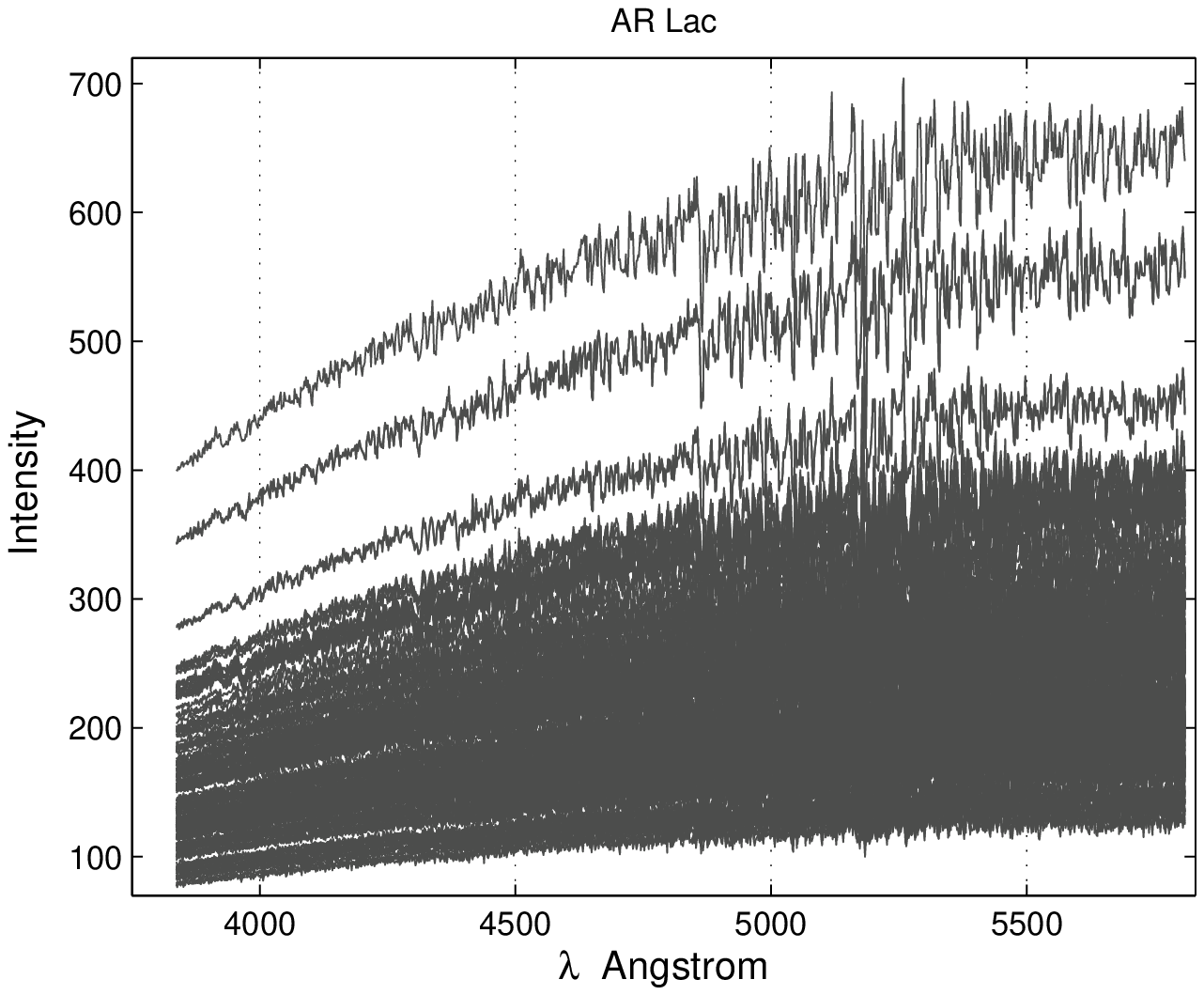,width = 1.15\linewidth} \caption{The set of spectra for AR Lac was taken on July 21-22, 2021 after correction for the response of the CCD detector.}\label{fig1}
\end{minipage}
\hfill
\begin{minipage}[t]{.45\linewidth}
\centering
\epsfig{file = 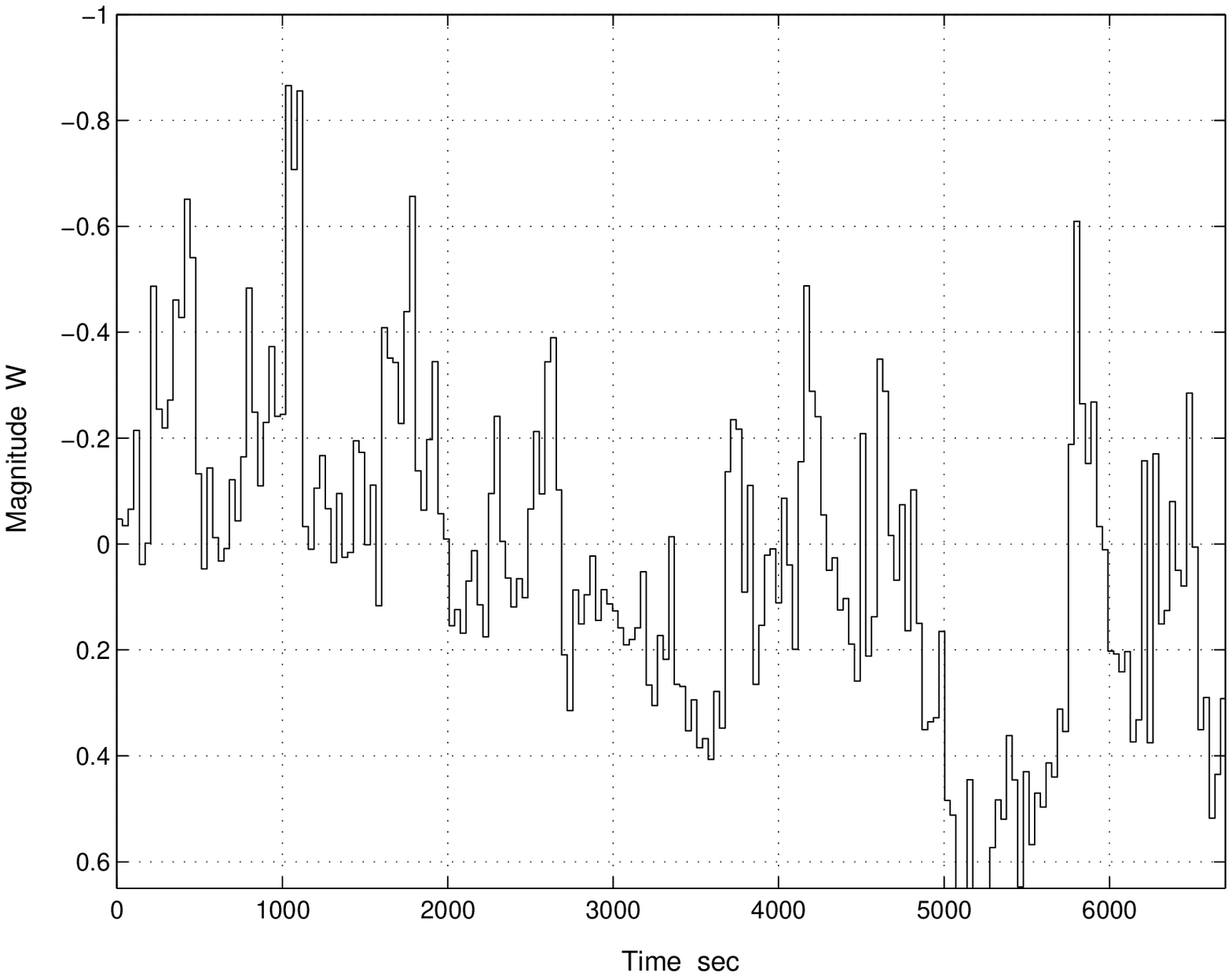,width = 1.0\linewidth} \caption{The white light curve for the July 21-22, 2021 flare on AR Lac.}\label{fig2}
\end{minipage}\end{figure}
%+++

%+++
\begin{figure}[!h]
\centering
\begin{minipage}[t]{.45\linewidth}
\centering
\epsfig{file = 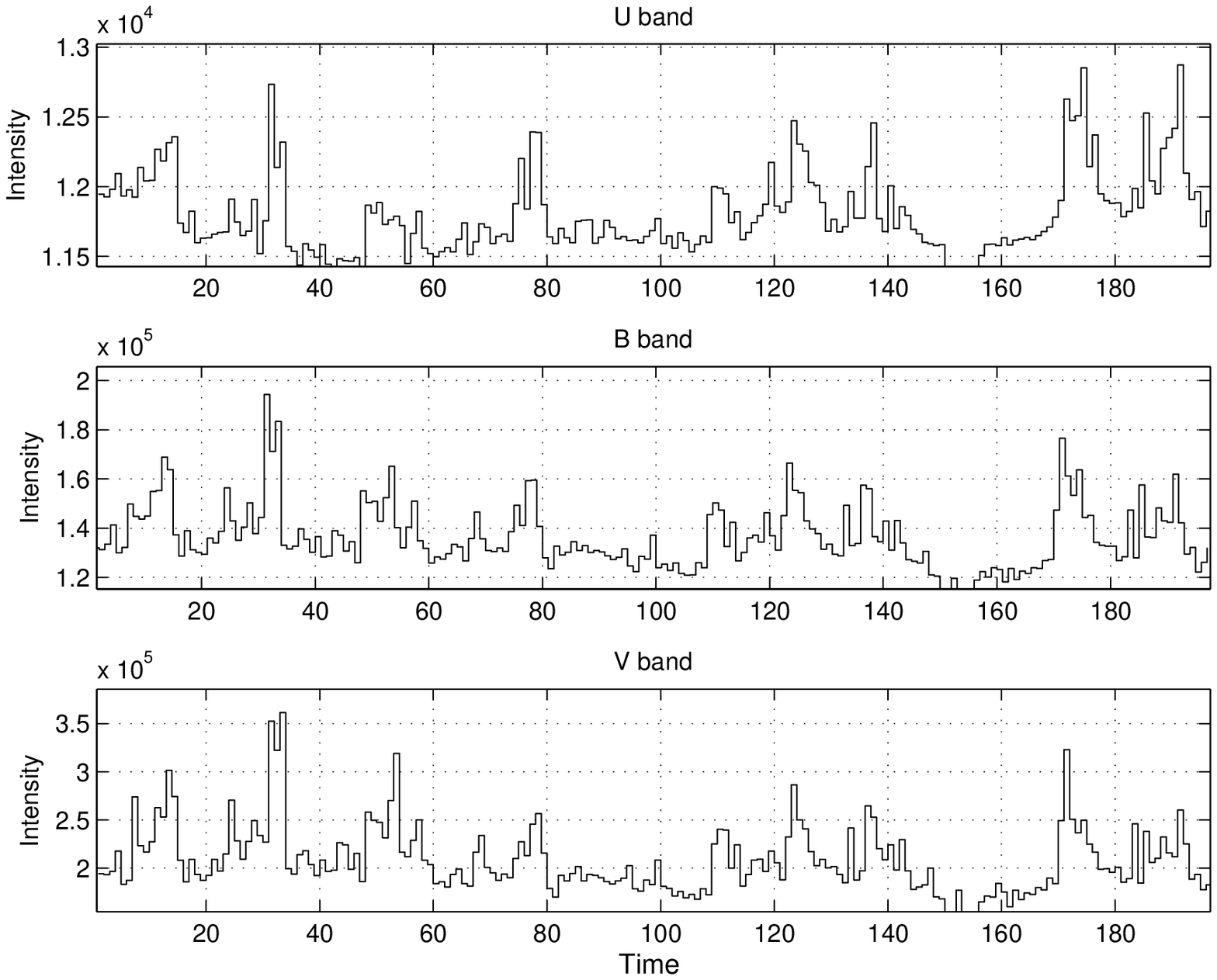,width = 1.0\linewidth} \caption{Graphs of radiation fluxes in filters $ U, B, V $}\label{fig1}
\end{minipage}
\hfill
\begin{minipage}[t]{.45\linewidth}
\centering
\epsfig{file = 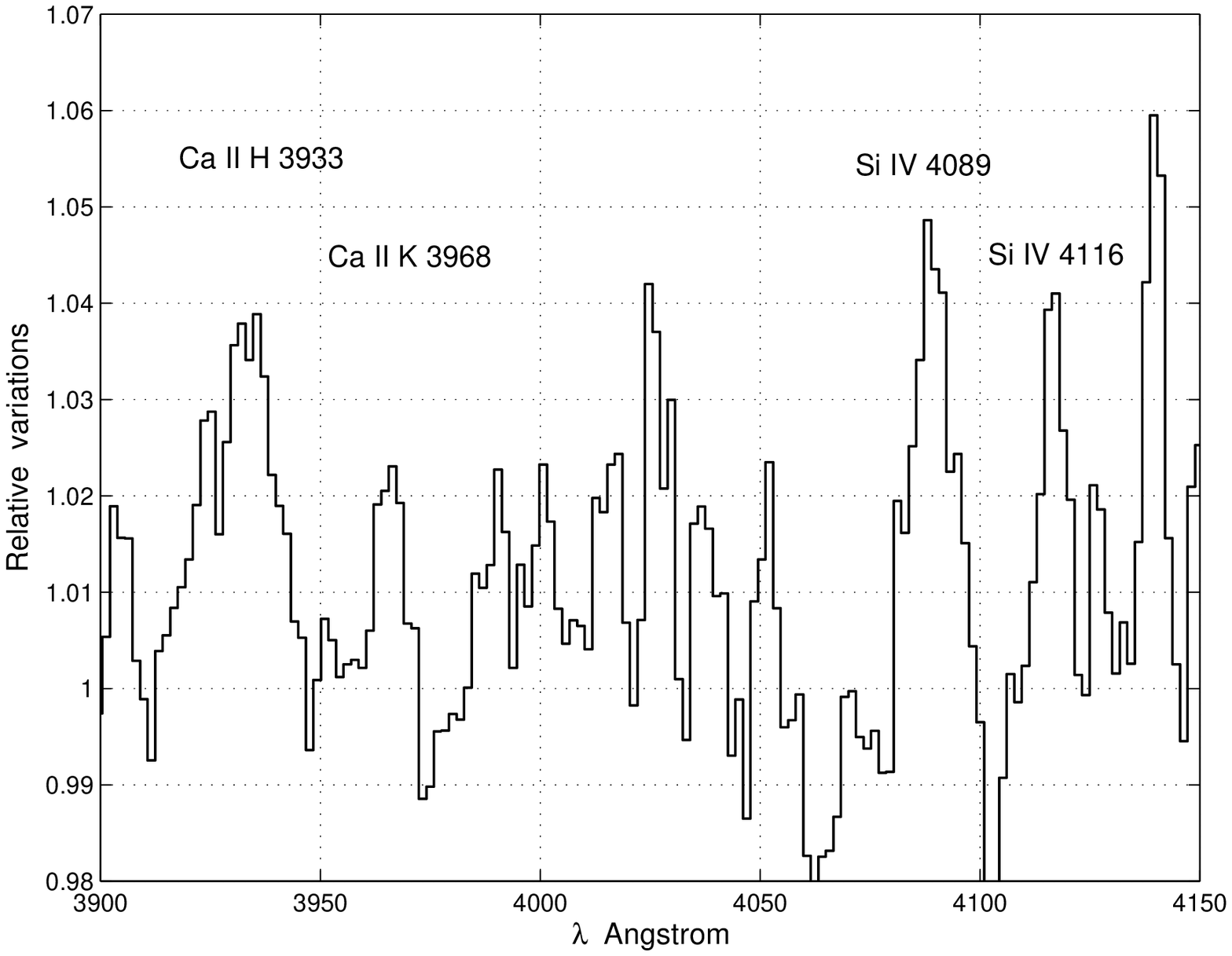,width = 1.0\linewidth} \caption{Relative variations in the spectrum of AR Lac.}\label{fig2}
\end{minipage}
\end{figure}
%+++

%+++
\begin{figure}[!h]
\centering
\begin{minipage}[t]{.45\linewidth}% ARLac_8z.eps
\centering
\epsfig{file = 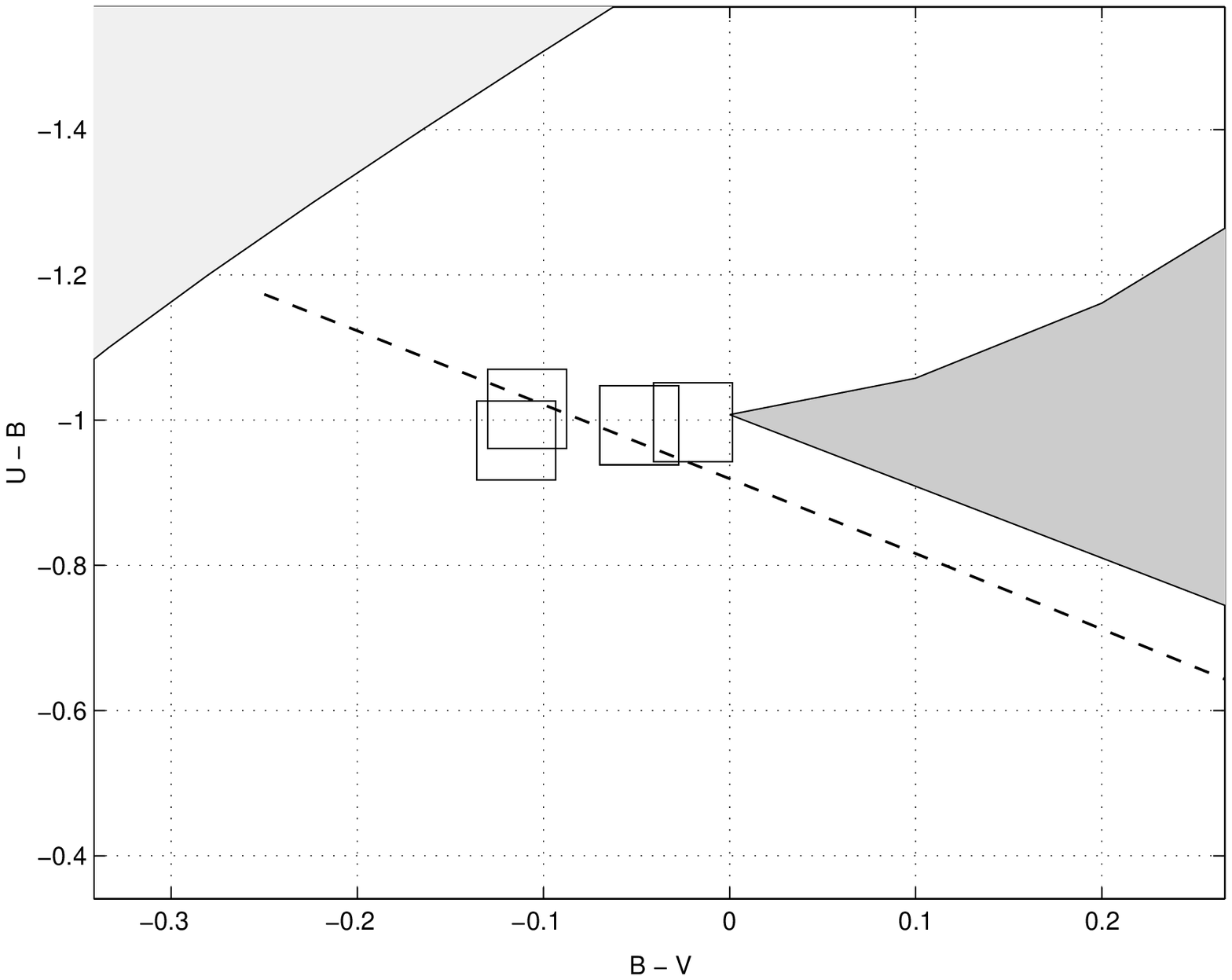,width = 1.0\linewidth} \caption{Color tracks of the July 21-22, 2021 flares on the (U - B)-(B - V) diagram. The squares mark flare maxima.}\label{fig5}% Color tracks of the July 21-22, 2021 flares on the (U - B)-(B - V) diagram.
\end{minipage}
\hfill
\begin{minipage}[t]{.45\linewidth}% ARLac_25.eps
\centering
\epsfig{file = 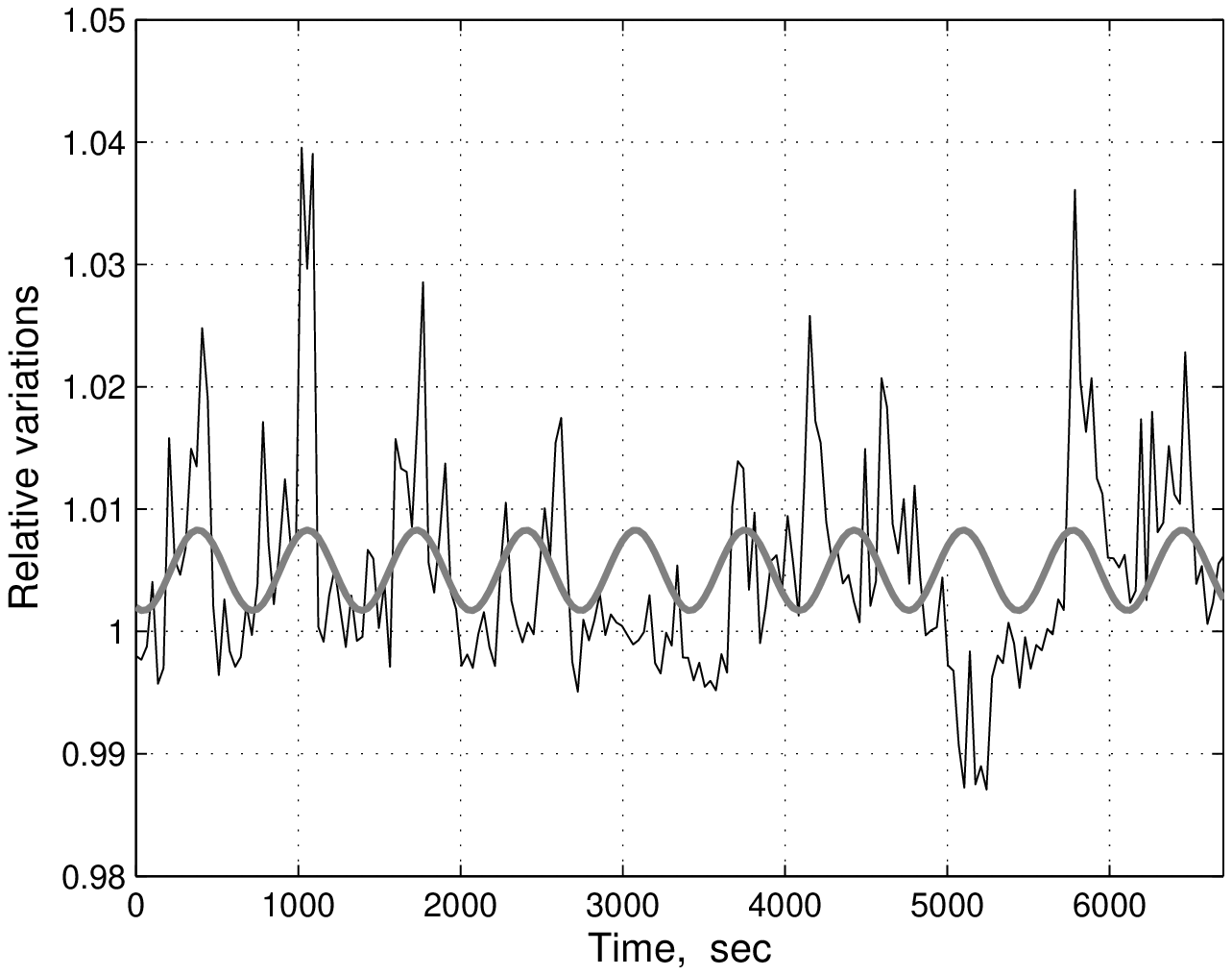,width = 1.15\linewidth} \caption{The detrended light curve of AR Lac in U-rays  with 11 min harmonic.}\label{fig62}% The detrended light curve of AR Lac with 11 min harmonic.
\end{minipage}
\end{figure}
%+++

\section*{\sc Flare colorimetry}
\indent \indent The modern approach to colorimetry is based on multicolor $UBVRI$ spectrophotometry in the spectral region from about of 3500 to approximately of 9000 \AA. Colorimetry is a quantitative method for analyzing radiation. It allows one to diagnose the radiation of a celestial body, determine the temperature, electron concentration, optical thickness of the emitting plasma based on theoretical diagnostic color diagrams calculated for various radiation sources.

At first, in each filter the intensities $FU_{0},FB_{0},FV_{0}$ in a quiet state of a star were calculated in Fig. 3. Then these values were subtracted from the corresponding values $FU,FB,FV$ of the flare intensities. Based on the obtained residues, the color indices of the intrinsic emission of the flare were determined:
\begin{equation}\label{}
    U-B=-2.5\lg\left[\frac{FU-FU_{0}}{FB-FB_{0}}\right]+\triangle_{UB}
\end{equation}
\begin{equation}\label{}
    B-V=-2.5\lg\left[\frac{FB-FB_{0}}{FV-FV_{0}}\right]+\triangle_{BV}
\end{equation}
where $\triangle_{UB}=0.26+2.5\lg\frac{FU_{0}}{FB_{0}}$, $\triangle_{BV}=0.72+2.5\lg\frac{FB_{0}}{FV_{0}}$ are normalization coefficients. These coefficients take into account the color indices of the star in a quiet state: $U-B$ = 0.26, $B-V$ = 0.72.

To develop two-color diagrams, we used the color indices of various radiation sources \cite{Staizys}, \cite{Chalenko}. The bright area of the diagram $(U-B)$ - $(B-V)$ in Fig. 5 corresponds to the color characteristics of a hydrogen plasma that is optically thin in the Balmer continuum with $T_{e} \sim $ 10000 K and $N_{e}$ from 10$^{14}$ to 10$^{10}$ cm$^{-3}$. The dark region of the diagram corresponds to an optically thick plasma with $T_{e}$ from 15000 to 8000 K. Absolute blackbody radiation is indicated by the bottom broken line.

Fig. 5 shows color tracks of the July 21-22, 2021 flares on the (U - B)-(B - V) diagram. The 95\% error squares are shown. The squares mark flare maxima. The color errors were calculated for Poissonian quantum fluxes.

The flare luminosity in the $U$ band can be determined by convolving the spectrum of a completely black body with the transmission curve of the $U$ filter. Flare area $s$ can be defined as \cite{Alekseev}:

\begin{equation}\label{}
    \frac{s}{S}=\left(10^{0.4 \triangle U}-1\right)\frac{FU_{0}}{FU}
\end{equation}
where $S$ is the area of the visible disk of the star, $\triangle U$ is the amplitude of the flare in the $U$ band, $FU_{0}$ and $FU$ are the Planck function of the photosphere and flare with effective temperature $T_{bb}$:
%\begin{equation}\label{}
%  F(T_{bb})=\int U(\lambda )/\lambda^{5}[\exp(1.4388/\lambda T_{bb} -1 ) ]d\lambda
%\end{equation}
\begin{equation}\label{}
  F(T_{bb})=\int \frac{U(\lambda )}{\lambda^{5}}\left[\exp\left(\frac{1.4388}{\lambda T_{bb}} -1 \right) \right]d\lambda
\end{equation}
Here $U (\lambda)$ is the transmission curve of the $U$ filter. Knowing the observed flare amplitude in the $U$ band, the photosphere temperature of AR Lac (5039 K, \cite{Joy}), and the temperature at the maximum of the flare, one can easily estimate the size of the flare using the above formulas. The linear size of the flares at the maximum luminosity is approximately 2\% of the radius of the star, or about 0.2\% of the area of the visible disk of the star.

The white light curve (Fig. 2) shows some periodicity in flashes. The detrended light curve of AR Lac (Fig. 6) shows a periodic "pilot signal", which control the flares of a star. Its amplitude is small, several thousandths of a magnitude in U-rays. We can see that all the flares appear at the highs of the pilot signal. This effect is a new phenomenon that is not taken into account in modern theories of stellar flares. For the first time periodic pilot signals have been detected in a quiet state of the flaring star YZ CMi \cite{Zhilyaev}.

\section*{\sc Conclusion}

Flares on an RS CVn type star AR Lac are an example of fast processes. The brightness of a star and the spectral composition of its radiation can change significantly in fractions of a minute.

We performed a fast AR Lac spectrophotometry on July 21-22, 2021 using the 2.0-m Carl Zeiss telescope at Terskol Observatory. The observations were obtained with a spectral resolution of R $\sim $ 1300; time resolution 34 s.

During the flare, additional emission was detected in the Ca II H\&K ($\lambda$ = 3933, 3968 \AA) and Si IV ($\lambda$ = 4089, 4116 \AA) lines. The excess of emissions is about two to five percent.

We estimated the UBV magnitudes of the flares by mathematically convolving the spectra with the filter transmission curves. UBV amplitudes reach 0.8 magnitudes.

From the color-color diagram (U-B)-(B-V) we found that the flares at maximum brightness radiate as a black body with a temperature of up to 12,000 $ \pm$ 300 K. The linear size of the flares is estimated at about 2\% of the stellar radius.

{}

%++++++++++

\end{document}